# Berry-Phase Breakdown and Semiclassical Reconciliation in Topological Dirac Fock-Darwin states

Ye-Ping Jiang

*Key Laboratory of Polar Materials and Devices (MOE) and Department of Electronics, East China Normal University, Shanghai 200241, China*

**Abstract**

I investigate the two-dimensional Dirac fermion analogue of artificial atoms (Fock-Darwin states, FD) in a circular n-p junction on a topological insulator surface. The FD states in this non-parabolic potential exhibit a unique electron-hole core-shell structure, where the strict Berry-phase switch (BPS) picture breaks down near criticality: the trapped electron-core states evolve into the envelope functions of quantized snake states. This contradicts the sharp BPS seen in experiments. Nevertheless, the BPS scenario remains valid when treating these envelope functions as effective confined states, thereby reconciling theory with experiment. The field-driven evolution from electrostatic to Landau-level confinement is tracked to 14 T experimentally and supported by simulations, establishing topological surface states as a tunable platform for Dirac physics beyond conventional quantum dots.

Email: ypjiang@clpm.ecnu.edu.cn

*Introduction*.—In conventional semiconductor quantum dots, electrostatically confined electrons form discrete energy levels known as Fock-Darwin (FD) states or "artificial atoms," which mimic atomic orbitals [1,2]. However, distinct phenomena emerge when such artificial atoms are realized in two-dimensional (2D) Dirac fermion (DF) systems [3-20]. A hallmark example is the Berry-phase switch (BPS), a unique magnetic-field response characteristic of confined DFs [7-10]. Experimentally, the confinement of 2D DFs can realized in circular n-p junctions (CNPJ) despite Klein tunneling [4-7,19-21], utilizing the ability of n-p junctions to efficiently reflect Dirac electrons with large incident angles [22-24].

In three-dimensional (3D) topological insulators (TI) [25,26], with their surfaces naturally host 2D DF systems, such CNPJs have recently been realized, manifesting as near-circular electron puddles induced by subsurface charged defects in slightly p-doped TI films [7]. BPS has been observed in these CNPJs. In TIs, the Fermi level can be tuned via growth-parameter control, and charged defects can be introduced during growth [27], offering a compelling platform for investigating exotic phenomena of confined DFs.

In this work, I carry out a detailed numerical investigation of FD states in such CNPJs, where the confining potential flattens away from the center. Unlike the commonly studied parabolic potentials [2,8,9], these realistic CNPJs host a unique electron-hole (e-h) core-shell configuration that becomes fully localized under an arbitrarily small magnetic field. I show that near the BPS critical point, the strict BPS picture breaks down: instead of vanishing, the confined core states lose their discrete identity and evolve into fully localized hybrid snake states that are delocalized across the core-shell interface due to strong Klein tunneling. This discrete-state picture appears to contradict the experimental observation of a well-defined BPS in such CNPJs [7]. I resolve this paradox by demonstrating that these hybrid snake states are in fact governed by well-defined envelope functions. The BPS concept remains fully valid when these envelope functions are treated as the effective confined states. Experimentally, I resolve these FD states and track their continuous evolution from the electrostatic-confinement regime into the Landau-level (LL) regime in a CNPJ fabricated on a seven-quintuple-layer $Sb_2Te_3$ film. The numerical envelope features agree well with the experimental data, as do the results of a semiclassical analysis based on a modified quantization condition that incorporates Klein tunneling. Furthermore, I identify two additional distinctive features of these Dirac FD states: (i) the pinning of the zeroth Landau level—composed of successive $(0, m)$ states—to the spatially varying Dirac-point energy, and (ii) a characteristic double-peak structure in the local density of states (LDOS) at the junction center, a direct signature of the Dirac spinor. Notably, the pinning behavior enables direct visualization of the junction's potential landscape by imaging the FD states at high magnetic fields. The extracted potential profile validates my reconciliation between the theoretical breakdown of the strict BPS picture and its experimental observation. My work establishes confined topological surface states as a versatile and tunable platform for exploring correlated Dirac fermion physics and phenomena beyond the reach of conventional semiconductor quantum dots.

*The electron core and hole shell structure of FD states.*—In my experiment, the CNPJ is a near-circular electron-puddle (~ 100 nm in width) induced by subsurface charged defects in the slightly p-doped 7-QL $Sb_2Te_3$ film [7]. The resulting confining potential follows a screened defect potential, $U(\boldsymbol{r}) = \mu_0 e^{-|\boldsymbol{r}|^2} + \mu_\infty$. Here, $\mu_0 + \mu_\infty$ and $\mu_\infty$ define the Dirac energy at the CNPJ center and in the p-doped background, respectively. As shown below, this potential profile leads to an e-h core-shell structure for the FD states.

In the presence of a magnetic field, the Dirac Hamiltonian for the surface states under confinement is $\varepsilon\psi(\boldsymbol{r}) = [v_F \boldsymbol{\sigma} \cdot \boldsymbol{p} + U(\boldsymbol{r})]\psi(\boldsymbol{r})$, where $\boldsymbol{p} = \boldsymbol{q} + e\boldsymbol{A}$. Here, $\boldsymbol{p}$, $\boldsymbol{q}$, $\boldsymbol{A}$ are canonical momentum, kinetic momentum, and magnetic vector potential, respectively. In the symmetric gauge and circular symmetric $U(\boldsymbol{r})$, $\boldsymbol{p}$ and $\boldsymbol{q}$ are given by $q_r = p_r = \pm\sqrt{(\varepsilon - U)^2 - (m/r - Br/2)^2}$ and $q_\theta = p_\theta - A_\theta = m/r - Br/2$, where $\varepsilon$, $U$, $r$, $B$, are in units of $\varepsilon^*$, $r^*$, $B^*$ as described in the caption of Fig. 1.

From the expression for $q_r$, the conditions $\varepsilon \geq E_1(r) = U + |m/r - Br/2|$ and $\varepsilon \leq E_2(r) = U - |m/r - Br/2|$ define the classically allowed regions for electron and hole states, respectively. Figures 1(a) and 1(b) illustrate these regions for $m = \pm 7/2$

states at 0 T, 0.87 T and 8 T. Here, $q_r = 0$ gives the classical turning points $r_e$, $r_e'$, $r_h$, $r_h'$ for electrons and holes trajectories. The second term under the square root in $q_r$, namely the kinetic angular momentum $q_\theta = m/r - Br/2$, determines the barrier width $d = r_h - r_e'$. This width increases with angular momentum *m* at zero field, analogous to the planar junction case [22]. This creates a classically forbidden gap (shadowed in Fig. 1) centered around the potential profile (the spatial dependent Dirac energy $E_D(r)$, dashed curve). This gap shrinks for states with smaller *m*, indicating that *m* acts as a tunneling barrier for trapped states. Unlike the commonly studied parabolic-like potentials (Fig. S1 of [28]), the FD states in the CNPJ defined by $U(\boldsymbol{r})$ consist of an electron core and a hole shell. These states become fully localized even in the presence of an arbitrary small magnetic field (Fig. 1(b)), thereby forming a distinct e-h core-shell structure.

*The semiclassical point of view of BPS.*—Furthermore, unlike states with $m < 0$, a touch point emerges between the electron-core and hole-shell for states with $m > 0$ (Fig. 1(b)). This critical point, denoted by $(\varepsilon_c, B_c)$, lies on the potential profile (the n-p boundary) and moves down in energy with increasing magnetic field. Here, $(\varepsilon_c, B_c)$ satisfies $B_c = 2m/\ln\,(\mu_0/\varepsilon_c - \mu_\infty)$ and is derived from the condition of $q_r = q_\theta = 0$. This point defines the critical condition for the BPS for both electron and hole states. Figure 1(c) shows various classical orbits ($m = 7/2$) in momentum space along the radial coordinate. For massless Dirac fermions, cyclotron orbits winding around the Dirac point acquire an additional $\pi$ Berry phase [7]. I find that the Berry phases for electron and hole states, $(\varphi_B^e, \varphi_B^h)_{E,B_c,m=7/2}$, are (0, 1) or (1, 0) (in units of $\pi$) for states below or above the critical energy, as depicted in Fig. 1(d). Moreover, $(\varphi_B^e, \varphi_B^h)_{\varepsilon_c,B,m=7/2}$ switch from (0, 1) at $B < B_c$ to (1, 0) at $B > B_c$ (Fig. S2 of [28]). In contrast, $(\varphi_B^e, \varphi_B^h)_{E,B,m=-7/2}$ remains (0, 0). Thus, from a semi-classical perspective, a $\pi$ Berry-phase switch occurs for confined state ($m > 0$) upon crossing the critical point. This switch alters the quantization condition and induces an energy jump of this state. Moreover, at this critical point $(\varepsilon_c, B_c)$, the classical electron trajectory approaches the n-p junction perpendicularly [see the critical condition for (0, 7/2) state in Fig. S2(b) of [28]], leading perfect transmission (Klein tunneling) and the formation of snake states (guided electron trajectories that propagate along an n-p junction in a perpendicular magnetic field) [29,30]. Note that the snake state touches the Dirac point in momentum space (orbit 6). Compared with those in the parabolic potential where the hole side trajectories of these states are open and approach infinity, the snake states in CNPJs here are bounded near the n-p interface and become quantized (orbits 3, 4, and 6; see the snake orbit of (0, 7/2) in Fig. S2(b) of [28]).

*Breaking down of BPS and the Reconciliation.*—Within the strict BPS picture, the FD states should disappear and reappear when approaching and departing from the critical condition, caused by the Klein tunneling between them and the hole continuum. This is exactly the case for parabolic potentials [8]. Thus, the strict BPS picture appears to break down near the critical point for FD states with such an e-h core structure, where they evolve into quantized snake states.

The numerical data shows the details of FD states near the criticality. Figure 2(a) displays the radial distribution of the local partial density of states (LPDOS, numerical as described in [28]) for $m$ = 7/2 at different magnetic fields. Here, $|U_1|^2 + |U_2|^2 = |U|^2$, where $U_1$, $U_2$ are the spin components of the Dirac wavefunction $\psi_m(r,\theta) = \frac{e^{im\theta}}{\sqrt{r}}\begin{pmatrix} U_1(r)e^{-i\theta/2} \\ U_2(r)e^{i\theta/2} \end{pmatrix}$. Appreciable LPDOS appears within the classically forbidden region, and the FD states become spectrally broadened near the critical point (0.87 T). A closer inspection at $r$ = 0.36 (Fig. 2(b), the solid curve) reveals that trapped ($n$, 7/2) states in the electron core lose their discrete identities near criticality, becoming instead the envelope functions of quantized snake states. This envelop nature is due to the strong coupling between the sparse electron-core levels and the dense hole-shell states. However, by reinterpreting the broadened spectral features as envelope functions of these snake states, I find that the peak energies still obey the BPS behavior. Notably, the horizontal lines, obtained semiclassically (described in the next section), match well with the peak centers of these envelope functions. Furthermore, the envelope functions broaden significantly for states approaching the critical point, consistent with a reduced lifetime due to Klein tunneling. Note that 0.87 T is the critical field of BPS for the (2, 7/2) state. The corresponding envelop peak disappears for this state (dashed line). This behavior is more clearly demonstrated in the field-dependent LPDOS (second derivative) shown in Fig. 2(c). As the system approaches the critical fields $B_c^-$ and $B_c^+$, the (2, 7/2) state maintains its identity as an envelope function, with its peak energy and width obeying the BPS behaviors. Here $B_c^-$ and $B_c^+$ denote the critical fields just before and after the BPS, respectively. Away from the criticality, the FD states become single-peaked. The trapped hole levels are shown in the LPDOS data at $r$ = 2. Consequently, the BPS scenario remains valid even near criticality when the envelopes are taken as the effective confined states.

Furthermore, Fig. 2(a) shows one distinct feature of Dirac FD states. The two spin components $U_1$ and $U_2$, of the FD states behave distinctly across the criticality. For a trapped electron state ($n$, $m$), the number of nodes for the component $U_2$ changes from $n+1$ to $n$, while that of $U_1$ remains $n+1$. Taking the (2, 7/2) state for example (dashed lines), the nodes for $U_2$ changes from 3 to 2, while that of $U_1$ remains 3. This situation is more clearly demonstrated in Fig. 2(d) where the spatial LPDOSs at the energy of this state below, at, and above the critical field are plotted. Particularly, for the (0, 7/2) state, the node for $U_2$ vanishes after BPS, indicating its fully spin-polarized nature. This applies to all (0, $m$) states that forming the zeroth LL after BPS.

*Semiclassical analysis of BPS.*—Despite the strong interplay between electron core and hole shell across the n-p junction, the energy of envelop peaks can be calculated semiclassically by taking into account a tunneling dependent reflection phase $\theta$ at the n-p junction. The processing orbits for electron in the center-symmetric field yields a two-valued momentum field $\boldsymbol{q}(\boldsymbol{r})$ (a two-torus, Fig. 3(a)) between the two classical return points $r_e$ and $r_e'$ [31]. Following EBK rule, in this case the semiclassical quantization condition for the trapped electron in the CNPJ can be considered along the radial and azimuthal coordinates $C_R$ and $C_\theta$ separately, $\oint_{C_R} p_r dr + \varphi_B^e|_R + \alpha + (-\frac{\pi}{2} + \theta) = 2\pi n$ and $\oint_{C_\theta} p_\theta dr + \varphi_B^e|_\theta = 2\pi m$, determining the radial and azimuthal quantum numbers $(n, m)$. For massless Dirac electrons on the TI surface, $\varphi_B^e|_\theta$ is always 1, yielding the half-integer value of $m$. The Berry's phase $\varphi_B^e|_R$, as has been discussed above, will switch from 0 to 1 at the critical condition [7]. Here $\alpha = -\frac{\pi}{2}$ is the reflection phase at the inner turning point $r_1$ [32]. The radial quantization condition is now reduced into

$$\oint_{C_R} p_r(r,\varepsilon,B)dr = 2\pi\left[n + \tfrac{1}{2}(1-\varphi_B^e|_R)\right] - \theta\ . \quad (1)$$

For the reflection phase $\theta$ at the n-p junction, I adopt the result of connection problem for a single n-p junction by WKB approximation [33,34], $\theta = Arg\left[\Gamma\left(1+i\frac{K}{\pi h}\right)\right] - \frac{\pi}{4} + \frac{K}{\pi h} - \frac{K}{\pi h}\ln\left(\frac{K}{\pi h}\right)$, where $K = \int_{r_i}^{r_o}|p_r^2|\,dr$. $\theta$ is thus a function of the tunneling probability $t = e^{-K/h}$ across the forbidden region from $r_e'$ to $r_h$ [14]. From Eqn. 1, the energy $\varepsilon_{nm}$ for each state ($n$, $m$) at different magnetic field $B$ can be obtained [28].

In Fig. 3(b), I compared the semiclassical results w/o the tunneling correction with the numerical ones for the $m = \pm 1/2$ states. The condition without tunneling (dashed) corresponds to the $\theta = 0$ case. The results with tunneling correction (solid) fit the numerical results much better, especially at low fields. Near the critical field at which Berry-phase switch happens, $\theta$ approaches $-\frac{\pi}{4}$ (dotted curves for (0, 1/2)), resulting in an upturn in energy to get larger radial actions compared with the $\theta = 0$ case. At fields well above the criticality $B_c$, $\theta$ approaches 0 and there is no difference between these two cases any more. In addition, each state with positive $m$ intersects with $B_c$ at two critical fields between which this state seems to disappear.

*The experimental field-dependent FD states*.—Figures 3(c) and 3(d) are the experimental field-dependent *dI/dV* spectra up to 8 T taken at $r = 0$ and 0.3 (in units of $r^*$) as well as the corresponding numerical data. The dashed line indicates the magnetic field of 1 T, below which the data shows BPS behavior as reported [7]. The FD states gradually evolve from the electrostatic-confinement region into the Landau region with the increasing magnetic field, with $N = n + \left(\left|m-\frac{1}{2}\right| - \left(m-\frac{1}{2}\right)\right)/2$ being the correspondence between the Landau level (LL) and FD indexes. For example, the 0-th LL is composed of the successive (0, $m$) states, with the energies increasing with $m$. The numerical results of the FD states match well with the experimental ones. In addition, the results obtained by my semiclassical analysis follow the same trend as the numerical data. The most distinct signature of Dirac fermions in the FD spectrum compared with Schrödinger electrons is the LDOS at the junction center (Fig. 3(c)). Each $n$-th LL is composed of double FD states, ($n$, +1/2) and ($n$-1, -1/2), which comes from the spinor nature of Dirac Hamiltonian. The only exception is the zeroth LL which only has one component, the (0, +1/2) state.

Note that an energy broadening of ~ 2 meV is used in the numerical data to match the experiments. Therefore, an energy broadening of about 2 meV is enough to smear out the enveloped quantized snake states to reconcile the disagreement between the experimental observed BPS and BPS-breaking down in the strict picture. One of the possible causes of the broadening may come from the scattering between surface states and the bulk states. Another cause may come from the broadening of trapped hole levels, possibly due to the irregular shape of the confining potential away from the center as indicated in the next section. Another point is that for the ($n$, 1/2) states whose BPS are most obvious [7], the quantization of snake states is less prominent (Figs. S3 and S4 of [28]).

*The spatial profile of FD states*.—The correspondence between the FD and the LL indexes can be seen more explicitly in the spatial profile of the confined states. In Fig. 4(a), the experimental data shows discrete FD states at various magnetic fields, most of which can be clearly indexed and in accord with the numerical data (Fig. 4(b)). The dotted curve in Fig. 4(b) denotes the (0, $m$) states, while the dashed one is the spatial profile of Dirac points. At 8 T, the (0, $m$) states are pinned at the Dirac energies, following a trend that mimics the potential profile (see also Fig. S5 of [28]). This is because the radial action (Eqn. 1) for the (0, $m$) state switches from $\pi$ - $\theta$ to - $\theta$ after the Berry-phase switch ($\varphi_B^e$ switches from 0 to π), where $\theta$ approaches 0 with the increasing magnetic field. A near-zero radial action means that the state resides nearly at the bottom of the electron-core (classically allowed region of the electron states), which is exactly the critical point (Dirac energy at the n-p interface). This can be seen clearly in the e-h core-shell structures at high fields in Fig. 1(b). Thus, from the spatial profile of the successive (0, $m$) states, the potential profile can be figured out. Figure 4(c) shows the *dI/dV* spectra at the center and at an off-center ($r$ = 0.3) position. The 0-th LL is composed of the (0, $m$ = 1/2, 3/2,…) states. The spatial distribution of these states thus the landscape of the potential profile can be visualized by taking the *dI/dV* mapping (Fig. 4(d)) at the energies of these successive peaks. The deformation of the potential from circular symmetry as displayed in Figs. 4(a) and 4(d) leads to the intrinsic broadening of the hole states that validates further the observed BPS.

At relatively high magnetic fields, compressible and incompressible regions form in a 2D electron system with potential variation because of the formation of LLs and the increased electron-electron interaction. The screened potential profile of the subsurface charged defects will be modified in this case. In the numerical data, without taking into account the electron-electron interaction, the potential profile (profiled by the (0, $m$ = 1/2, 3/2,…) states) remains unchanged. Experimentally, the potential profile at 14 T is steeper by 8 meV than that at 8 T. In addition, there are signatures of kinks (arrowed in Fig. 4(a)) in the potential profile, which indicate the reduced screening as well as the formation of incompressible regions at relatively high magnetic fields. Compared with the graphene case [11], the kinks are very weak and no charging effect can be observed probably because of the strong screening from the bulk states or there are relaxation channels to the bulk states for the localized LL electrons.

*Conclusions*.—I have analyzed FD states in a CNPJ of TI surface states with a confining potential typical of screened subsurface charged defects. Unlike FD states of Dirac systems in different potentials, the trapped electrons and holes in this realistic potential exhibit a unique e-h core-shell structure that can be completely confined in an arbitrary field. The confined electron-core states become envelope functions of the quantized snake states due to strong Klein tunneling. By taking the envelope peaks as the effective confined states, I establish an effective model that reconciles the strict picture with experiments. This model successfully matches the experimental data of FD states up to 14 T, thereby directly visualizing distinct properties of Dirac fermions and the magnetic-field-dependent potential landscapes. The unique core-shell structure of CNPJs may enable highly tunable coupling between core and shell states when arranged into CNPJ arrays. This work demonstrates that the surface of a 3D TI, when tuned near charge neutrality [35], provides an ideal platform for exploring rich physics in 2D Dirac fermion systems, such as electron optics and quantum information technologies.

## ACKNOWLEDGEMENT

The author thanks one of the anonymous referees for insightful comments. The author acknowledges the supporting from National Natural Science Foundation, Ministry of Science and Technology of China (Grants No. 12474478, 2022YFA1403100, 92065102).

## REFRENCES

**Figure Captions**

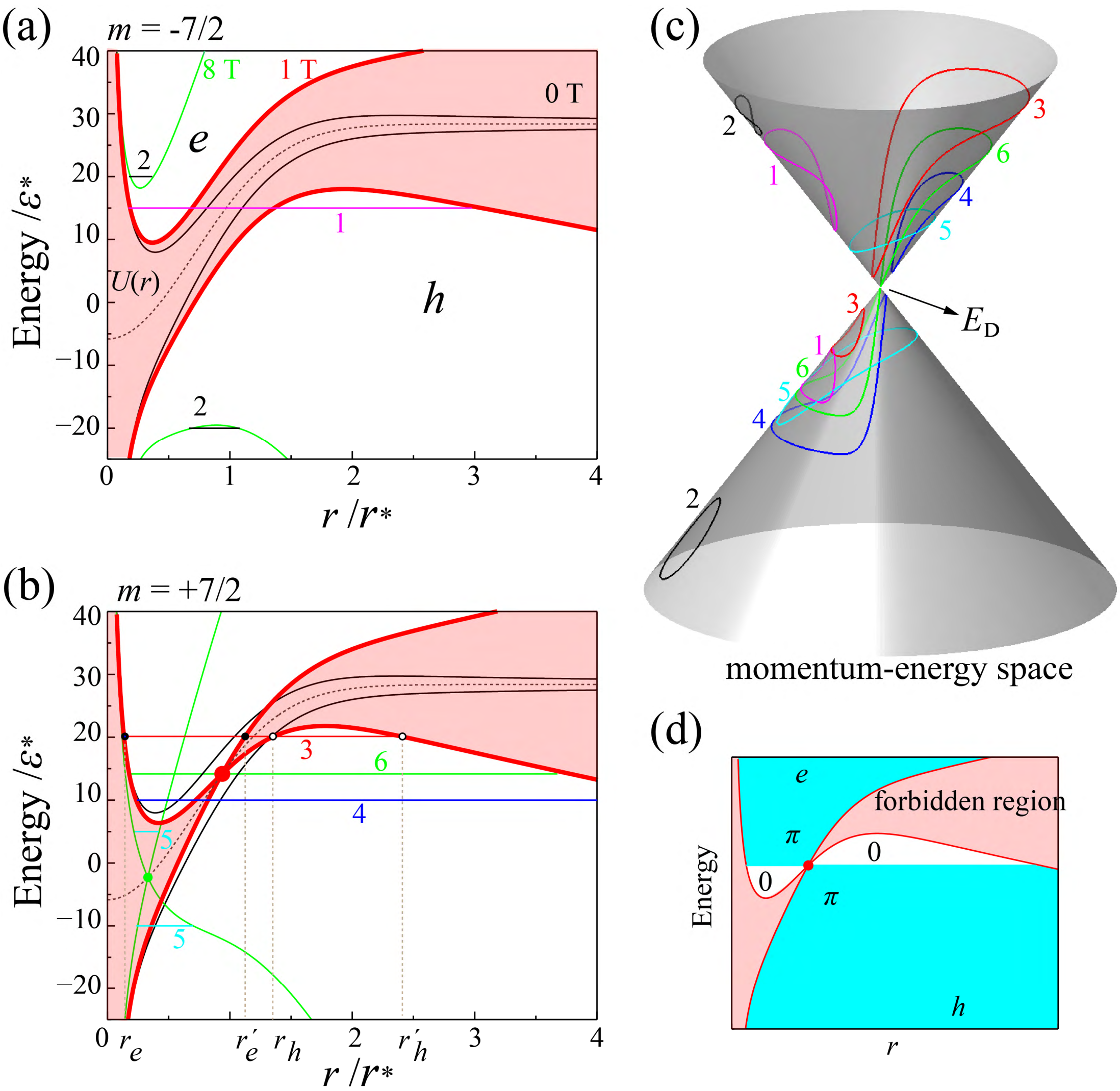

FIG. 1 (color online). The core-shell struture of FD states. (a) (b) The classical allowed regions at 0 , 1 and 8 T for $m = \pm 7/2$, respectively. Here the shadowed are the forbidden regions at 1 T. The dashed curves are the potential profiles ($E_D(r)$). Here $r_e$, $r_e'$ and $r_h$, $r_h'$ define the classical allowed regions at $E = 20$ for electrons and holes, respectively. The solid dots are the critical points. (c) The classic orbits in the momentum-energy space along the radial coordinate (numbered correspondingly in (a) and (b)) at different energies and magnetic fields. (d) The diagram $\varphi_B(\varepsilon, r)$of Berry's phase at a specific magnetic field. In the simulation, we use a radial length scale $R_0$ = 50 nm, a Fermi velocity $v_F \sim 4.3 \times 10^5$ m/s$^2$. Here, $\mu_0 = -34.2\varepsilon^*$, $\mu_\infty = 28.4\varepsilon^*$, energy scale $\varepsilon^* = \hbar v_F/R_0 \sim 3.91$ meV. I use the convention that the states with positive $m$ show Berry-phase switch in positive $B$.

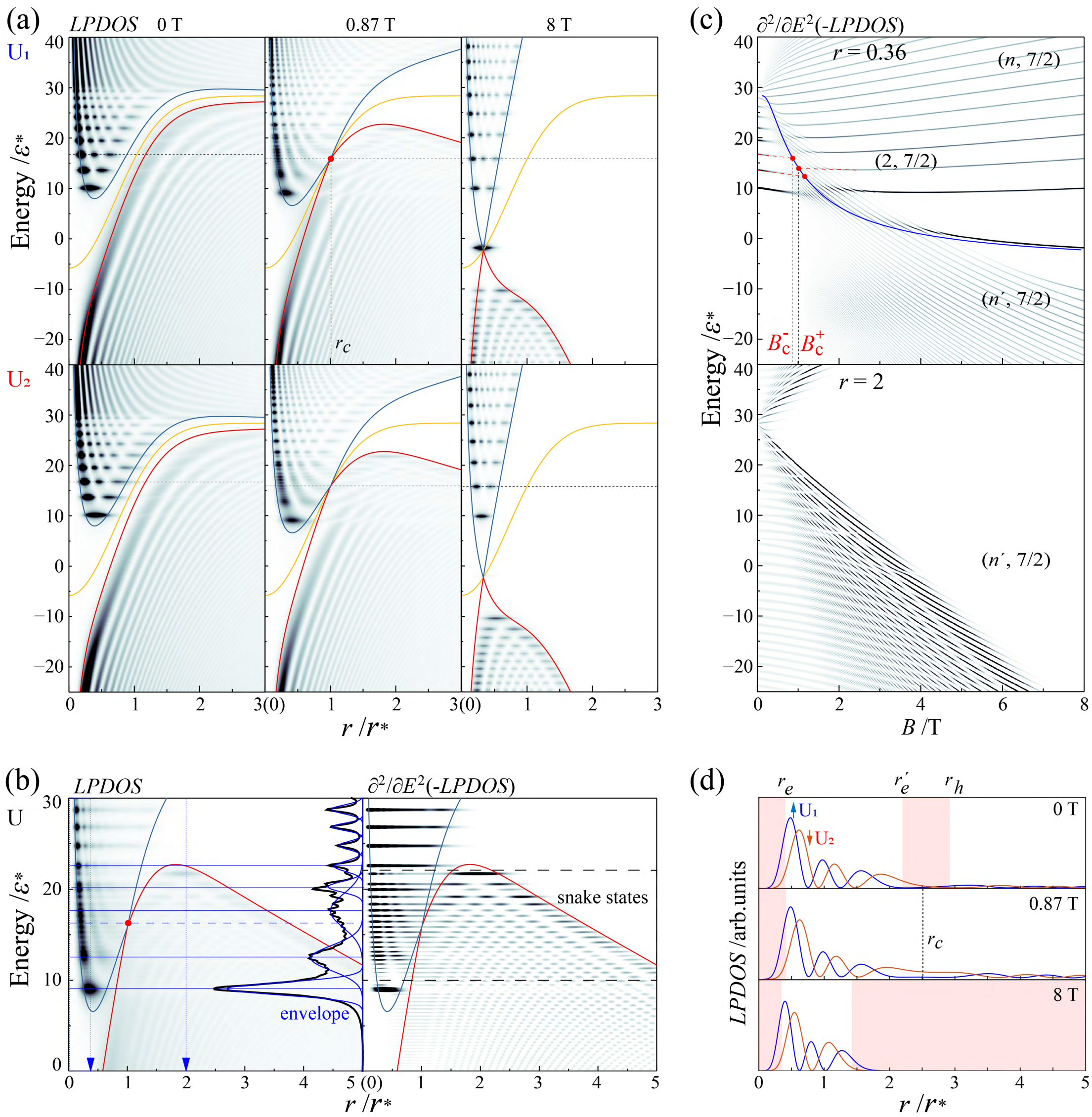

FIG. 2 (color online). The detailed nature of Berry-phase switch. (a) The spatial and energy dependent LPDOS for $m$ = 7/2 (spin-resolved, numerical) at 0 T, 0.87 T and 8 T, respectively. The classical allowed regions are also plotted. The energies of (2, 7/2) are indicated by the dashed horizontal lines. (b) LPDOS and its second derivative near the critical energy at 0.87 T. The solid curve in the top-middle figure is the LPDOS at $r$ = 0.36. (c) The field-dependent LPDOS (second derivative) of the $m$ = 7/2 states at $r$ = 0.36 and 2, respectively. The solid curve is $B_c$ for $m$ = 7/2. (d) The two spin components of LPDOS of the (2, 7/2) state along $C_R$. The shadowed are the classical forbidden regions.

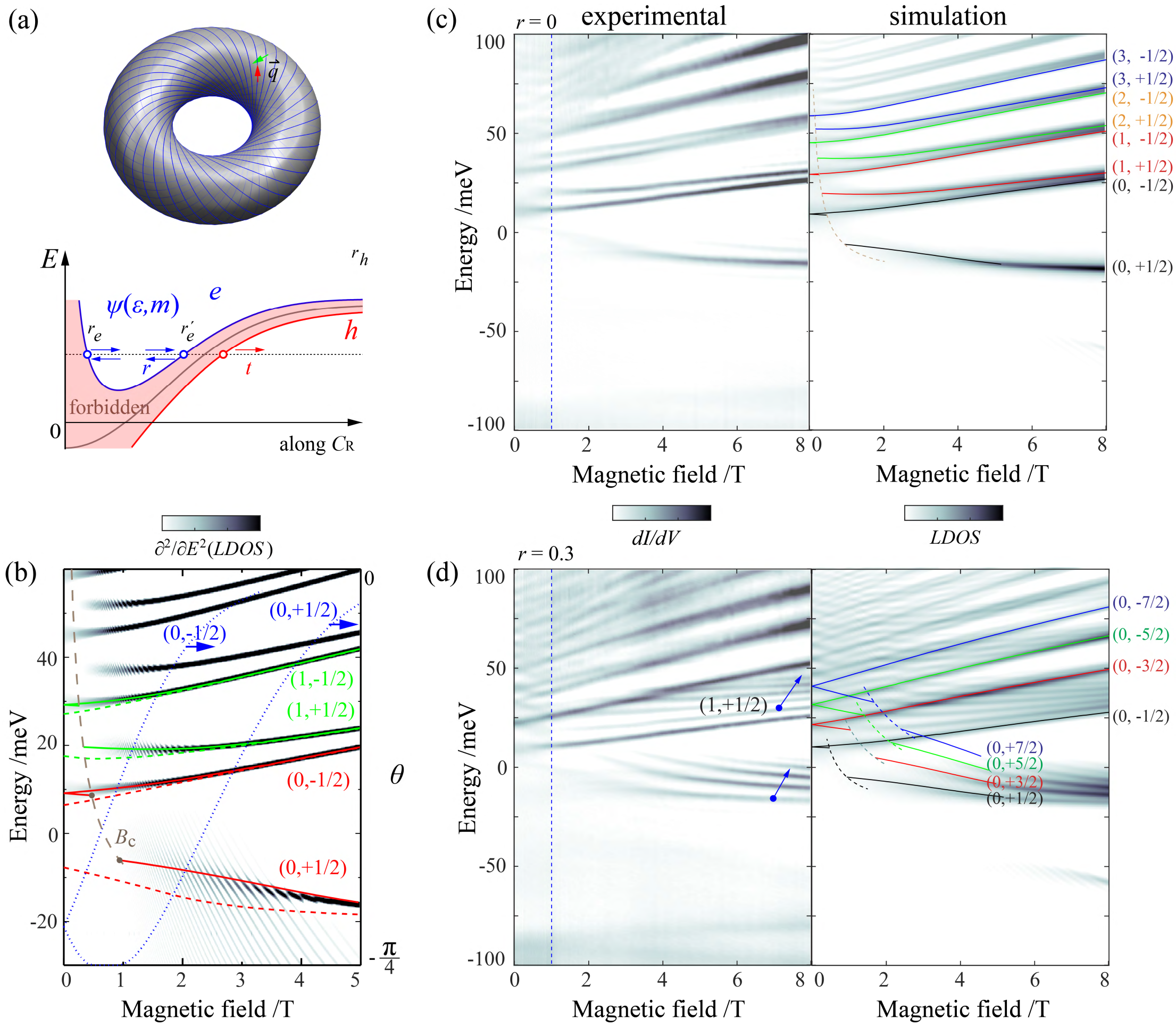

FIG. 3 (color online). Field-dependent FD states obtained semiclassically, numerically and experimentally. (a) The vector field for the electron states of (0, 7/2) at 0 T as well as the diagram of tunneling between the electron core and hole shell at the n-p juction. (b) The comparison between the FD states (0, ± 1/2), (1, ± 1/2) obtained numerically (second derivative, colored map) and semiclassically. Here we show both the semiclassical results with (solid curves) and without (dashed curves) the tunneling correction. The blue dotted curves are the $\theta$ ($B$) for (0, ± 1/2). (c) (d) The expimental FD states obtained at the center and an off-center position in the CNPJ as well as the corresponding numerical data. The dashed lines indicate the region of [0, 1] T where Berry-phase switch behaviors can be observed [7]. The semiclassical data (solid curves) is plotted along with FD states obtained numerically. The dashed curves are $B_c(\varepsilon)$ for the states with different $m$.

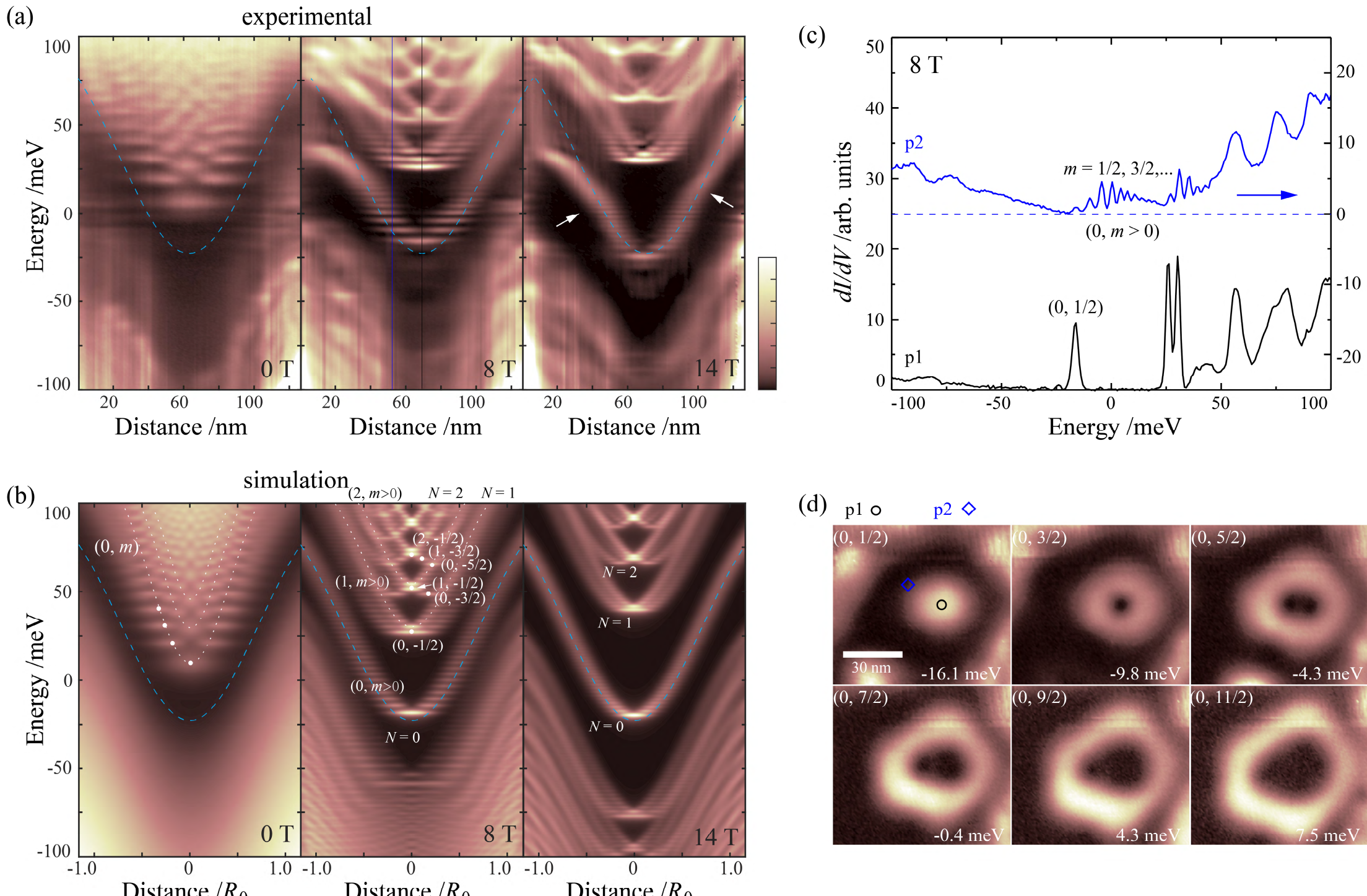

FIG. 4 (color online). The potential profile revealed by the (0, *m*) states. (a) (b) The spatially resolved *dI/dV* and the corresponding numerical results long a line (126 data points, 125 nm) across the CNPJ at 0 T, 8 T and 14 T, respectively. The dotted curves in the simulation at 0 T indicates the (0, *m*), (1, *m*) and (2, *m*) states. The dashed curves is the potential profile. (c) The *dI/dV* taken at two different positions in the CNPJ (indicated by the dotted lines in (a)) at 8 T. (d) The *dI/dV* mapping taken at energies corresponding to those of the (0, *m* > 0) states (peaks in (c) around 0 meV) at 8 T.